\documentstyle[aps,prl,epsf]{revtex}

\newcommand{\bx}[0]{{\bf x}}

\begin{document}
\twocolumn[\hsize\textwidth\columnwidth\hsize\csname @twocolumnfalse\endcsname
\title{On Nonlinear Diffusion with Multiplicative Noise}
\author{Miguel A. Mu{\~{n}}oz$^{(1,2)}$ and Terence Hwa$^{(1)}$}
\address{$^{(1)}$Department of Physics, University of California at San Diego,
La~Jolla, CA 92093-0319}
\address{$^{(2)}$Dipartamento di Fisica,
 Universit\'a di Roma ``La Sapienza'',~P.le
A. Moro 2, I-00185 Roma, Italy}
\date{\today }
\maketitle

\begin{abstract}

Nonlinear diffusion is studied in the presence of multiplicative noise. 
The nonlinearity can be viewed as a ``wall'' 
limiting the motion of the diffusing field. A dynamic phase transition occurs
when the system ``unbinds'' from the wall. Two different universality classes, 
corresponding to the cases of an ``upper'' and a ``lower'' wall, are
identified and their critical properties are characterized. While the 
lower wall problem can be understood by applying the knowledge of 
linear diffusion with multiplicative noise, the upper wall problem exhibits an
anomaly due to nontrivial dynamics in the vicinity of the wall.
Broad power-law distribution is obtained throughout the bound phase.

\vspace{10pt}

\vspace{10pt}

\end{abstract}

]


The diffusion equation with multiplicative noise has been used as a
paradigm to describe a large class of nonequilibrium dynamical 
processes~\cite{general} ranging from light propagation in random 
media to wealth fluctuations in economic systems. 
In this paper, we study the effect of nonlinearities, which inevitably
occur in such processes due to the existence of characteristic
scales of the diffusing fields. For instance, the growth of a population
is bounded from above by a characteristic density determined by the
competition of resources~\cite{pop}, and an investor's wealth can be taken as
bounded from below by a fixed income source~\cite{LS0}. At sufficiently 
coarse grained scales, these effects may be
captured by the following {\em nonlinear} diffusion equation with
multiplicative noise, 
\begin{equation}
{\partial }_tn={\bf \nabla }^2n+a\,n-b\,n^{1+p}+n({\bf x},t)\,\eta ({\bf x}%
,t)  \label{lan0}
\end{equation}
where $a$ and $b$ are constants, $p$ specifies the degree of nonlinearity,
and $\eta ({\bf x},t)$ is a 
Gaussian noise with zero mean and the variance $\left\langle \eta ({\bf x}%
,t)\eta ({\bf x}^{\prime },t^{\prime })\right\rangle =2D\,\delta ({\bf x}-%
{\bf x}^{\prime })\,\delta (t-t^{\prime })$. Note that Eq.~(\ref{lan0})
should be interpreted either in the Ito or Stratanovich sense~\cite{VK},
depending on the details of the specific process being considered. 
We shall fix the nonlinearity coefficient to be $b=p$ from here on, 
so that for $p>0$, the nonlinear term acts like a soft {\em upper wall} 
preventing $n(\bx,t)$ from reaching large values even if $a$ is large and 
positive, while for $p<0$, the nonlinear term acts like a soft 
{\em lower wall} repelling $n(\bx,t)$ from approaching zero even 
for large and negative $a$'s. The specific choice of $p$ depends on 
the symmetries and constraints of the system. 
The limits $p\rightarrow \pm\infty $ mimic the effect of {\em hard%
} upper/lower walls  and are also of interest.

It will be convenient to cast the upper/lower wall problems in a more
``symmetric'' form. By making use of the Cole-Hopf transform, $h(\bx,t)=
\log n(\bx,t)$, we have
\begin{equation}
\partial _th=\mbox{\boldmath $\nabla$}^2h+(\mbox{\boldmath $\nabla$}%
h)^2+(a-a_0)-pe^{ph}+\eta(\bx,t) ,  \label{kpz}
\end{equation}
which is a Langevin equation similar to the KPZ equation describing the
kinetic roughening of a growing interface $h$~\cite{KPZ,KS}, with
an additional drift term $a-a_0$ ($a_0=0$ for Stratanovich dynamics and $%
a_0=D$ for Ito dynamics~\cite{VK}), 
and an ``exponential wall'' at $h\approx 0$. Thus
a wall in the diffusion problem corresponds to a wall also in the interface
problem. Eq.~(\ref{kpz}) shows that the sign of $p$ determines the {\em %
orientation} of the wall while the magnitude of $p$ describes the hardness
of the wall. In the interface representation, it is clear that for large
positive [negative] $a$'s, the system is pushed against the upper [lower]
wall, while for large negative [positive] $a$'s, the system is pushed away
from the wall to $h=-\infty $ [$h=+\infty $], corresponding to $n=0$ [$%
n=\infty $]. A critical point separates the two regimes where the system
``unbinds'' from the wall. In this paper, we study the nature of this {\em %
dynamic unbinding transition} for walls characterized by different $p$'s. We
will show that while the soft and hard walls yield the same critical
phenomena, the upper and lower walls yield two {\em different} universality
classes.


We start with a review of the exactly solvable
``zero-dimensional'' (single-site) version of Eqs.~(\ref{lan0}) and (\ref
{kpz})~\cite{GS}. Without spatial couplings, the stationary probability distribution $%
\widetilde{P}(h)$ of Eq.~(\ref{kpz}) is simply 
\begin{equation}
\widetilde{P}(h)\propto \exp \left[ \frac{a-a_0}Dh\right] \cdot \exp \left[ -%
\frac 1De^{ph}\right] ,  \label{Ph.0}
\end{equation}
where the second exponential merely enforces the constraint of the wall,
suppressing $\widetilde{P}$ for $h>0$ [$h<0$] if $p>0$ [$p<0$].
From (\ref{Ph.0}), the
stationary distribution of $n$ is easily obtained using the relation $%
P(n)=n^{-1}\,\widetilde{P}(\log n)$, yielding $P(n)\propto
n^{[(a-a_0)/D]-1}\exp (-n^p/D)$. Note that
the distribution functions are normalizable as long as $(a-a_0)\cdotp > 0$,
and hence a power-law distribution in $n$ is obtained for a large range
of parameter values.  [A similar result was obtained previously
for the case of a hard lower wall~\cite{LS,SC}; 
it was proposed as an explanation of the
observed power-law  distribution of wealth~\cite{wealth}.] 

As $a\to a_0^{\pm }$, the average $h$ diverges according to Eq.~(\ref{Ph.0}) 
as $\langle h\rangle \sim \mp |a-a_0|^{-1}$ for the upper/lower wall, 
indicating the onset of an ``unbinding transition'' of $h$ from the wall. 
This transition can be characterized quantitatively by singularities in 
moments of $n$. From $P(n)$, we find for the upper wall problem
$\langle n^m\rangle \sim (a-a_0)^{{\beta }_m}$ and ${\beta }_m=1$ for
all integer $m$'s.  The distribution collapses towards $%
P(n)=\delta (n)$ as $a\to a_0^{+}$. For the lower wall, $\langle n^m\rangle $
{\em diverge} as $a\rightarrow a_0^{-}$ since $P(n)$ has no upper cutoff. It
is convenient to characterize the phase transition in this case by monitoring 
$\bar{n}\equiv 1/n$.
One finds $\langle \bar{n}^m\rangle \sim (a_0-a)^{\bar{\beta}_m}$, with $%
\bar{\beta}_m=1$ also. The symmetry between the upper and lower wall
problems is evident from the $h\rightarrow -h$ symmetry of Eq.~(\ref{kpz}) 
in the absence of spatial couplings. Note that the asymptotic
critical properties given above are independent of the parameter $p$,
indicating that in the vicinity of the unbinding transition, where $\langle
n\rangle $ approaches zero or infinity (or as $\langle h\rangle \to \mp
\infty $), the detailed form of the wall at finite $n$ (or at $h\approx 0$)
is {\em irrelevant}.

Time-dependent properties of the system can be obtained by solving the full
Fokker-Planck equation~\cite{VK,GS}.
Qualitative features can be obtained alternatively by considering
the simpler Langevin equation for $h(t)$. We illustrate the solution by
analyzing the problem with a lower wall $(p<0)$. For $a<a_0$, $h$ is
confined to the range $0\lesssim h\lesssim \delta h$, where $\delta h\sim
(a_0-a)^{-1}$ is the scale of typical fluctuations in $h$. Right at the
critical point $a=a_0$, $\delta h$ diverges, i.e., $\delta h(t)\rightarrow
\infty $ as $t\rightarrow \infty $. The form of $\delta h(t)$ is dictated by
the equation of motion at the critical point, $\partial _th=-p\,e^{ph}+\eta
(t).$ $\delta h(t)$ must be {\em at least} of $O(t^{1/2})$ due to the
random forcing $\eta $; the presence of the wall can only speed up the
motion away from the wall. On the other hand, as $h$ drifts far away from
the wall, it should not be affected by the wall. Thus $\delta h(t)\sim
t^{1/2}$ at the transition, with the distribution given by the scaling form 
\begin{equation}
\widetilde{P}(h,t)=\left( \delta h\right) ^{-1}\,g\left[ h/\delta
h(t)\right] \quad {\rm for}\quad ph<0.  \label{ph0}
\end{equation}
In (\ref{ph0}), the factor $\left( \delta h(t)\right) ^{-1}$ provides the
proper normalization, 
and $g(y)$ is a scaling function with the limiting behaviors $g(y)\approx {\rm const}$
for $|y|\ll 1$ and $g(y)\rightarrow 0$ for $|y|\gg 1$. 
The distribution $P(n,t)$ follows from (\ref{ph0}), yielding $\langle
n^m\rangle \sim \delta h\sim t^{1/2}$ for all positive moments $m$.


We now proceed to characterize the behavior of the system in arbitrary
spatial dimension $d$. This is accomplished by generalizing the above
zero-dimensional picture and utilizing the known properties of the KPZ
equation in finite dimensions~\cite{KPZ}. 
The Cole-Hopf transformation has already been exploited
in Ref.~\cite{Yuhai} to derive some properties of the system with a soft
upper wall $(p>0)$. Some of the arguments there (but not all \cite{beta})
can be generalized also to the case of a lower wall.

Close to the critical point, $h(\bx,t)$ is on average far from the
wall, and the main source of fluctuation comes from the stochastic KPZ
equation without walls (for all length scales below the correlation length $%
\xi $). The scaling properties of the KPZ equation can be described in terms
of the dynamic exponent $z$ alone~\cite{KPZ}: 
the typical correlation time is $\tau
\sim \xi ^z$ and typical height fluctuation is $\delta h\sim \xi ^{2-z}$.
These properties yield the scaling form for the additive renormalization $%
\langle (\mbox{\boldmath $\nabla$}h)^2\rangle $ of the bare drift $(a-a_0)$,
with $\langle (\mbox{\boldmath $\nabla$}h)^2\rangle =a_1-a_2(\xi )$, where $%
a_1\propto D$ is a constant and $a_2(\xi )\sim (\delta h/\xi )^2\sim \xi
^{-1/(2z-2)}$. Thus one obtains the important result $\xi \sim |a-a_c|^{-\nu
}$ with $a_c=a_0-a_1$~\cite{note} and $\nu =1/(2z-2)$ for both 
the upper and lower wall. 
In particular, in one dimension where it is known exactly that $z=3/2$%
, we have $\nu =1$, while in $d=2$, $z\approx 1.60$ yields $\nu \approx 0.83$%
. The exponents $\nu $ and $z$ can be used to relate the exponents $\beta $ [%
$\overline{\beta }$] and $\theta $[$\overline{\theta }$] which describe the
behavior of the ``order parameter'' $n$ [$\overline{n}=n^{-1}$] close to the
critical point for the problem with an upper [lower] wall. From the
definitions $\left\langle n^m\right\rangle \sim \left| a-a_c\right| ^{\beta
_m}$ and $\left\langle n^m\right\rangle \sim t^{-\theta _m}$ at $a=a_c$, it
follows that $\theta _m=\beta _m/(\nu z)$; similarly, $\overline{\theta }_m=%
\overline{\beta }_m/(\nu z)$.

To find the exponents $\beta$ and $\theta$, we first perform a naive
scaling analysis. Let us assume
that the probability distribution $\widetilde{P}(h,t)$ at the critical point
 is still given by the
scaling form (\ref{ph0}) as in the zero-dimensional case, but with $\delta
h(\tau )\sim \tau ^\omega $ where $\omega =(2-z)/z$ in $d>0$. We then
find $\langle n^m \rangle_{{\rm upper}} \sim \langle \bar{n}^m\rangle_{{\rm %
lower}} \sim \delta h$ as before, yielding $\theta_m=\overline{\theta }_m=\omega
=(2-z)/z$ and $\beta _m=\overline{\beta}_m=(2-z)/(2z-2)$ for all $m$'s.
These exponents take on the value $\theta=1/3$, $\beta =1/2$ in 1d and $%
\theta \approx 0.25$ and $\beta \approx 0.33$ in 2d.


In the above analysis, we have not distinguished between systems with 
a lower or upper wall. On the other hand, it is evident that the equation of
motion (\ref{kpz}) for $p>0$ and $p<0$ are different in finite $d$
due to the presence of the nonlinear term $(\mbox{\boldmath
$\nabla$}h)^2$. Thus the problems with upper and lower wall can no longer be
mapped onto each other. In order to see whether the orientation of the wall
is relevant to the critical behaviors, and whether the above scaling ansatz
is correct, we perform a numerical simulation in one spatial dimension. We
expose now the algorithm and the main results.

To speed up the computations, we use a discrete model which, in the absence
of walls, belongs to the KPZ universality class~\cite{KS}. 
This model is similar to that of ballistic deposition: At each
point $x$ of a one dimensional lattice of size $L$, we define a continuous
height variable, $h_t(x)$. At every time step a new variable $h^{\prime
}(x)=h_t(x)+a+\eta _t(x)$ is obtained, with $\eta _t(x)$ being a random
number uniformly distributed in the interval $[0,1]$. The value of $h$ is
updated simultaneously according to the rule $h_{t+1}(x)=\max \left[
h^{\prime }(x\pm 1)+\gamma ,\,h^{\prime }(x,t)\right] $, where $\gamma =0.1$
is a fixed constant, and periodic boundary condition is applied. The
parameter $a$ controls the average drift of the system and is the only
tunning parameter in this model. A {\em hard wall} at $h=0$ is introduced 
by including the additional rule $h_t(x)=\min \left[ h_t(x),0\right] $ 
for an upper wall or $h_t(x)=\max \left[ h_t(x),0\right] $ for a lower wall. 

We describe first the results for an upper wall. The first step to studying
critical properties is to locate the critical point accurately. Starting
with the initial condition, $h_0(x)=0$ for all $x$, we let the system evolve
long enough so that a stationary state is reached ($\sim 10^5$ time steps are
typically required for system size of $L \sim 10^3$). To compute any
magnitude we average over up to 1000 independent runs. For a given size $L$,
the critical point $a_c(L)$ is taken as the first value of $a$ for which the
steady state value $\left\langle n\right\rangle =\left\langle \exp
(h)\right\rangle $ becomes indistinguishable from zero as $a$ is decreased.
A plot of $a_c(L)$ vs $1/L$ is shown in Fig.~1(a). Linearity of the
data suggests $\nu \approx 1.$ Extrapolating to $L\rightarrow \infty $, we
get $a_c\approx -1.5750$. The scaling of $\left\langle n\right\rangle $ and $%
\left\langle n^2\right\rangle $ upon approaching the critical point $a_c(L)$
are shown in Fig.~1(b) for $L=400$. We obtain the exponents $\beta _1\approx
\beta _2=1.50\pm 0.15$. In Fig.~1(b), we plot $\langle n(t)\rangle $ and $%
\langle n^2(t)\rangle $ vs. $t$ at $a=a_c$, and find $\theta _1\approx \theta
_2=1.10\pm 0.12$.

\begin{figure}
\epsfysize 2.0in
\centerline{\epsfbox{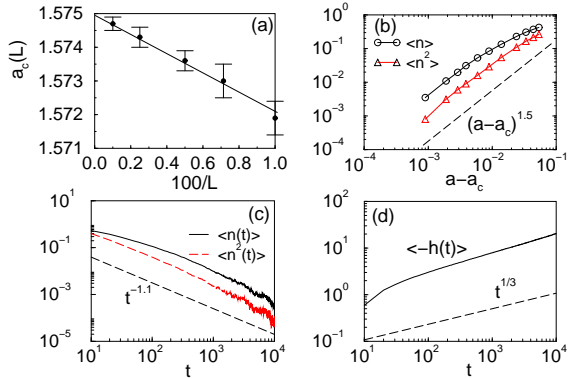}}
\vspace{10pt}
\caption{ Results with an upper wall: (a) Size dependence of the critical 
point $a_c(L)$; (b) first two moments of the spatially-averaged $n$
vs. $a-a_c$ in the steady state; (c) time dependence of $\langle n \rangle$
and $\langle n^2 \rangle$ at $a=a_c$; (d) $\langle -h(t)\rangle$ at $a=a_c$.}
\end{figure}

The scaling of $\langle h\rangle $ at $a=a_c$ 
is shown in Fig.~1(d). We
find the exponent $\omega \approx 0.33$. To compute the dynamic exponent
$z$, we perform a slightly different simulation~\cite{Gras}: We take as
initial condition a state in which $h$ is a large negative constant at every
lattice point except for the origin where $h=0$. The spreading of this
``localized seed'' is followed, and from the time evolution of the average
size of the ``infected region'', we find $z=1.50\pm 0.05$.
All the measured exponents coincide, within numerical accuracy, with those 
computed previously~\cite{Yuhai} for Eq.~(\ref{lan0}) with $p=1,2$, 
i.e., soft walls. This indicates that the introduction of an upper
soft or a hard wall to a model belonging to the KPZ universality class yields
the same critical behavior. As found in Ref.~\cite{Yuhai}, the exponents $%
\nu $ and $z$ agree with the result of the scaling analysis. Also the
scaling relation $\theta _m=\beta _m/(\nu z)$ \cite{Yuhai,Gras} is
satisfied. However, the values of $\beta $ and $\theta $ are significantly
different from those derived above using the naive scaling analysis.

We next describe the results for a lower wall. Following the same analysis
procedure for the ``order parameter'' $\left\langle \bar{n}\right\rangle
=\left\langle \exp \left( -h\right) \right\rangle $, we obtain the critical
point at $a_c\approx -1.5743$, and $\nu \approx 1$ (see Fig. 2(a)).
From Fig.~2(b), we find $\bar{\beta}_1\approx \bar{\beta}_2=0.48\pm 0.05$ 
\cite{beta} and from Figs.~2(c) and 2(d), we find $\bar{\theta}_1\approx 
\bar{\theta}_2=0.32\pm 0.03$, $\omega =0.33\pm 0.04$. The dynamical exponent 
$z\approx 1.5$ is again determined using the seed-spreading method (a suitable initial
condition is $h(x)=0$ for all $x$ except for the origin where $h(0)>0$).
Note that in contrast to the case with an upper wall, all of the measured
exponents here are consistent with the expected results based on the naive
scaling analysis, including the values of $\bar{\beta}$ and $\bar{\theta}$.
The clear difference between $\beta ,\theta $ and $\bar{\beta},\bar{\theta}$
indicate that {\em problems with upper and lower walls belong to different
universality classes}. 
\begin{figure}
\epsfysize 2.0in
\centerline{\epsfbox{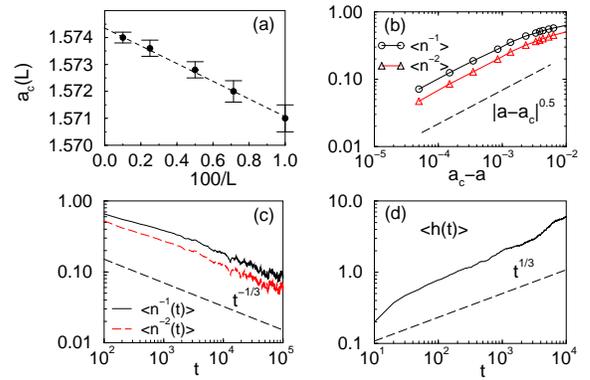}}
\vspace{10pt}
\caption{ Results with a lower wall: (a) Size dependence of the critical 
point $a_c(L)$; (b) first two moments of the spatially-averaged $\bar{n}$
vs. $a_c-a$ in the steady state; (c) time dependence of $\langle \bar{n} 
\rangle$ and $\langle \bar{n}^2 \rangle$ at $a=a_c$; 
(d) $\langle h(t)\rangle$ at $a=a_c$.}
\end{figure}

As the fluctuations in $\langle h(t) \rangle$ obey the same scaling law 
($\omega \approx 0.33$) for both the upper and lower wall problems 
(see Figs.~1(d) and 2(d)), 
differences in the scaling of $\langle n^m\rangle _{{\rm upper}}$ 
and $\langle \bar{n}^m\rangle _{{\rm lower}}$ must result from
differences in the shape of the distribution $\widetilde{P}(h,t)$, or more
specifically, in the form of the scaling functions $g(h/t^\omega )$ defined
in Eq.~(\ref{ph0}). Our numerical results suggest that the
zero-dimensional behavior of $g(y)\approx {\rm const}$ for $|y|\lesssim 1$
and $g(y)\rightarrow 0$ for $y\gg 1$ is obtained in 1d for the lower wall
problem only. The numerically obtained forms of the scaling function $%
g(h/t^{1/3})$ for the upper and lower wall are presented in Fig.~3(a) and
Fig.~3(b) respectively. Note that for large $|h|/t^{1/3}$, the distribution
function drops off sharply (approximately exponentially) for both cases.
Thus the assumed form for the scaling function $g(y)$ is
satisfied for large $y$. However, the distributions are qualitatively
different for small values of $|h|/t^{1/3}$, reflecting qualitative differences
in the interaction between $h$ and the upper/lower wall at $h= 0$.

\begin{figure}
\epsfysize 1.5in
\centerline{\epsfbox{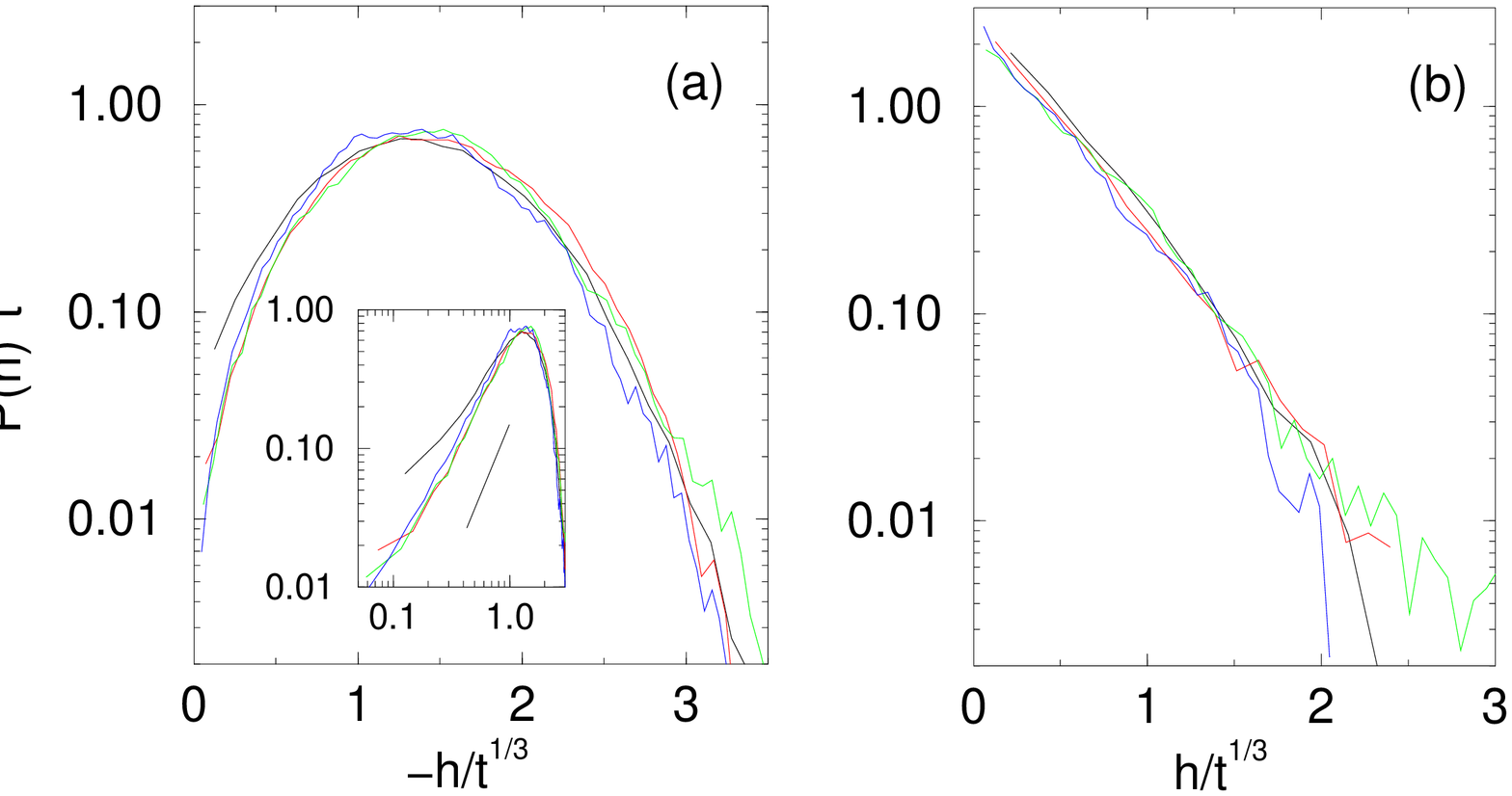}}
\vspace{10pt}
\caption{Scaled probability distribution
 $\widetilde{P}(h,t)\cdot t^{1/3}$ for
10 samples of size $L=3000$,
 with (a) an upper wall [inset: log-log plot; straight
line indicates $(|h|/t^{1/3})^2$], and
(b) a lower wall. The different
curves are for $t= 500, 2500, 5000, 10000$.}

\end{figure}

To appreciate the origin of this difference, let us consider the evolution
of a piece of {\em flat} interface $h=0$ at the critical point $a=a_c=a_0-a_1
$. The local drift rate  $\langle \partial _th\rangle \approx \langle (%
\mbox{\boldmath $\nabla$}h)^2\rangle -a_1$ is {\em negative} since $h=0$ and $%
\mbox{\boldmath $\nabla$}h=0$ there. In the case of a lower wall, this
negative drift has no effect since the interface cannot penetrate below the
wall. A steady state results from a dynamic balance between the driving
force $(a-a_c<0)$ which ``flattens'' pieces of the interface against the
wall, and the noise $\eta $ which roughens the interface. Thus, the situation
is very similar to the zero-dimensional case, and we expect the exponent
results obtained above from the naive scaling analysis to be valid for all $d$.

 The situation is
very different for the case of an upper wall, since the negative drift
represents a significant force {\em repelling} the interface from the wall.
The effective repulsion exerted by the upper wall is in fact {\em long ranged%
}: Suppose the interface is on average at a distance $\bar{h}$ from the
wall. Then interfacial roughness is ``interrupted'' by the wall at a scale $%
\bar{\xi}\sim \bar{h}^2$ in 1d, leading to an effective repulsive force $%
\langle \partial _th\rangle \sim -1/\bar{\xi}^\nu\sim -1/\bar{h}^2$.
Thus, a steady state can only be achieved by keeping the interface {\em away}
from the wall. 
This necessarily leads to the {\em depletion} of $\widetilde{P}(h,t)$ for
small $|h|$. In analogy to the problem of a random walker in the vicinity of
an absorbing wall, we expect $\widetilde{P}$ to have a power law form, $%
\widetilde{P}(h,t)\sim (|h|/t^{1/3})^\sigma $ for $|h|\ll t^{1/3}$. Our
numerical results (Fig.~3(a) inset) are consistent with this power law form,
with the exponent $\sigma \approx 2$.
The new form of $\widetilde{P}$ for the upper wall 
allows us to obtain moments of $n$ in terms of the exponent $\sigma $. 
We find $\theta _m=(1+\sigma )\cdot\omega \approx 1$. 
Our numerical result $\theta _m\approx 1.1 \pm 0.1$ (Fig.~1(c))
 is consistent with this exponent relation. 
The anomalous exponents can in principle be computed by analyzing the
effect of nonlinearity along the line of Ref.~\cite{Yuhai}, but starting
with the exact knowledge of the KPZ equation in $d=1$.
 
Away from the critical point, the
form of $\widetilde{P}(h)$ is simply obtained by replacing $\delta h(t)\sim
t^\omega $ with $|a-a_c|^{-\nu (2-z)}=|a-a_c|^{-1/2}$ in $d=1$. Our numerics
for the lower wall problem (Fig.~3(b)) suggests that $\widetilde{P}%
(h)\propto \exp \left[ -c\,h(a_c-a)^{1/2}\right] $ where $c$ is a
nonuniversal constant. This exponential distribution of $h$ (which we have
also checked directly by numerics) leads to a power law distribution of $n$
as in the zero-dimensional case. Again, we expect this result to be valid
for all $d$. For the upper wall problem, the
depletion effect described by the extra factor $[|h|(a-a_c)^{1/2}]^\sigma $ in 
$\widetilde{P}(h)$ leads to logarithmic corrections to the
power law form of $P(n)$.

In summary, nonlinear diffusion with
multiplicative noise has been analyzed in terms of the {\em dynamic
unbinding} of an interface from a wall. Two distinct
universality classes are obtained for the upper and lower walls
in finite spatial dimensions. The upper wall problem,
which describes various saturation effects,  exhibits anomalous scaling
due to nontrivial interaction with the wall. In contrast, 
the lower wall problem, which models the interaction
of a growing interface with a substrate and the fluctuation of wealth in
economic systems, can be understood 
by combining and applying the knowledge of the 0-d problem
and the KPZ equation.  It is interesting
to note that the lower wall problem, in particular the numerical
algorithm used here for the hard lower wall, resembles the
algorithms used for the ``local, gapped'' alignment of DNA
sequences~\cite{align,HL}. (Similarly, the 0-d problem 
is analogous to {\em gapless} local alignment~\cite{karlin}.)
Divergent critical fluctuations described in this work
have a significant effect on the optimization of sequence alignment
and will be discussed in detail elsewhere.

We are grateful to M.~L\"assig, Yuhai~Tu, G.~Grinstein and A.V.~Gruzinov 
for discussions.
M.A.M.~acknowledges financial support from the University of Granada and
the hospitality of the Physics Department at UC San Diego.
T.H.~is supported by an A.P. Sloan Research Fellowship
and an ONR Young Investigator Award. 


\begin{references}
\bibitem{beta}  Note that the calculation used in \cite{Yuhai} to derive the
bound $\beta>1$ for the $p=1$ case cannot be extended to $p<0$.
\bibitem{note} Note that the critical value $a_c=a_0 - a_1 \propto D$
can be either negative (Stratanovich dynamics) or positive (Ito dynamics).
Consequently, a system with $a=0$ may be on either side of the transition
depending on the details of the dynamics. Similar dependence may exist
and become relevant in the magnetic dynamo problem~\cite{dynamo}.
\end{references}
\end{document}